# Observation of multiple types of topological fermions in PdBiSe

B. Q. Lv,[1,2,†] Z.-L. Feng,[1,3,†] J.-Z. Zhao,[4,5,†] Noah F. Q. Yuan,[2] A. Zong,[2] K. F. Luo,[6] R. Yu,[6] Y.-B. Huang,[7] V. N. Strocov,[8] A. Chikina,[8] A. A. Soluyanov,[9,10] N. Gedik,[2] Y.-G. Shi,[1*] T. Qian,[1,11*] H. Ding[1,11*]

[1] *Beijing National Laboratory for Condensed Matter Physics and Institute of Physics, Chinese Academy of Sciences, Beijing 100190, China*
[2] *Department of Physics, Massachusetts Institute of Technology, Cambridge, MA 02139, USA*
[3] *University of Chinese Academy of Sciences, Beijing 100049, China*
[4] *Co-Innovation Center for New Energetic Materials, Southwest University of Science and Technology*
[5] *Theoretical Physics and Station Q Zürich, ETH Zürich, 8093 Zürich, Switzerland*
[6] *School of Physics and Technology, Wuhan University, Wuhan 430072, China*
[7] *Shanghai Synchrotron Radiation Facility, Shanghai Institute of Applied Physics, Chinese Academy of Sciences, Shanghai 201204, China*
[8] *Paul Scherrer Institute, Swiss Light Source, CH-5232 Villigen PSI, Switzerland*
[9] *Physik-Institut, University of Zürich, Winterthurerstrasse 190, CH-8057 Zürich, Switzerland*
[10] *Department of Physics, St. Petersburg State University, St. Petersburg, 199034, Russia*
[11] *CAS Center for Excellence in Topological Quantum Computation, University of Chinese Academy of Sciences, Beijing 100190, China*

[†] These authors contributed to this work equally.

[*] Corresponding authors: dingh@iphy.ac.cn, tqian@iphy.ac.cn, ygshi@iphy.ac.cn

## Abstract

**Topological semimetals with different types of band crossings provide a rich platform to realize novel fermionic excitations, known as topological fermions. In particular, some fermionic excitations can be direct analogues of elementary particles in quantum field theory when both obey the same laws of physics in the low-energy limit. Examples include Dirac and Weyl fermions, whose solid-state realizations have provided new insights into long-sought phenomena in high-energy physics. Recently, theorists predicted new types of fermionic excitations in condensed-matter systems without any high-energy counterpart, and their existence is protected by crystalline symmetries. By studying the topology of the electronic structure in PdBiSe using density functional theory calculations and bulk-sensitive soft X-ray angle-resolved photoemission spectroscopy, we demonstrate a coexistence of four different types of topological fermions: Weyl, Rarita-Schwinger-Weyl, double class-II three-component, and charge-2 fourfold fermions. Our discovery provides a remarkable platform to realize multiple novel fermions in a single solid, charting the way forward to studies of their potentially exotic properties as well as their interplay.**

Topological semimetals are characterized by symmetry-protected band crossings, which may give rise to quasi-particle excitations and topological features that underlie exotic transport and optical properties [1]. The most famous examples are Dirac [2-9] and Weyl semimetals [10-18]. The Dirac semimetal has two doubly-degenerate bands that cross linearly at isolated points, where the Dirac point has a topological charge of $C = 0$ [Fig. 1(a)(iii)]. By contrast, a Weyl semimetal features non-degenerate band crossings, where the crossing point (Weyl point) carries a topological charge of $C = \pm 1$ [Fig. 1(a)(i)]. Both systems host low-energy excitations analogous to those in high-energy physics and are described by the $4 \times 4$ Dirac equation [19] and the $2 \times 2$ Weyl equation [20], respectively.

Dirac and Weyl semimetal states have been realized and well studied in various compounds in condensed-matter physics [2-18], which provides an alternative and simple platform to study the novel physics of elementary particles, especially the long-sought Weyl fermions, in high-energy physics. Only three types of fermions are allowed in high-energy physics – Dirac, Weyl, and Majorana fermions, while the zoology of topological fermions in semimetals is much richer. This is because low-energy excitations in topological semimetals are constrained by the space group symmetries of the crystal, which are usually much lower than the Poincaré symmetry imposed by the quantum field theory. Indeed, various types of semimetals with topological fermions have been proposed in the past few years [2-4,10-14,21-30], which can be briefly classified by the degeneracy and topological charge of band-crossing points in the momentum space [Fig. 1(a)]. Panel (i) shows three types of Weyl fermions with twofold degeneracy. In type-I Weyl semimetals, the Weyl points arise from two linear dispersions with opposite Fermi velocities along all momentum directions, whereas in type-II Weyl semimetals Fermi velocities share the same sign along a certain direction [14]. Besides the type-I and type-II fermions with $C = \pm 1$, theorists predicted another type of Weyl fermion, quadratic Weyl fermion with $C = \pm 2$, which is formed by two quadratic bands [11]. Panel (ii) shows symmetry-protected crossings with threefold degeneracy, giving rise to two classes of unconventional three-component fermions: class-I is characterized by one non-degenerate and one doubly-degenerate linear band crossings at separate points, which carry no topological charge [25-27]; class-II is formed by three non-degenerate bands with $C = \pm 2$ [23]. Moving on to fourfold band degeneracy in panel (iii), besides the well-known Dirac fermions, we note two types of unconventional fermions differentiated by their topological charges: charge-2 fourfold fermions ($C = \pm 2$) and Rarita-Schwinger-Weyl fermions ($C = \pm 4$) [29]. For even higher band degeneracies, shown in panels (iv) and (v), theory predicted double class-II three-component fermions ($C = \pm 4$) and double Dirac fermions ($C = 0$) [22,23], which can be viewed as a nontrivial doubling of class-II three-component fermions and Dirac fermions, respectively. Despite many proposals of material candidates for hosting these unconventional topological fermions [21-30], experimental evidence has been scant. To date, only class-I three-component fermions have been observed [31,32].

In this Letter, we expand the experimental horizon by simultaneously observing four different types of symmetry-stabilized topological fermions in a single solid-state system, PdBiSe. Using angle-resolved photoemission spectroscopy (ARPES) operating at the soft X-ray energy, we identified the coexistence of Weyl ($C = \pm 1$), Rarita-Schwinger-Weyl ($C = + 4$), double class-II three-component ($C = - 4$), and charge-2 fourfold fermions ($C = + 2$) at different time-reversal invariant momenta of PdBiSe, all matching our expectation from density functional theory (DFT) calculations.

High-quality single crystals of PdBiSe were grown by the self-flux method. Soft X-ray ARPES measurements were performed at the Advanced Resonant Spectroscopies (ADRESS) beamline at the Swiss Light Source (SLS) with a SPECS analyser [33], and at the 'Dreamline' beamline of the Shanghai Synchrotron Radiation Facility (SSRF) with a Scienta Omicron DA30L analyser. Most of the soft X-ray ARPES data were taken at the ADRESS beamline with the photon energy ranged from 400 eV to 1,000 eV. The combined (beamline and analyser) experimental energy resolution of the soft X-ray ARPES measurements varied from 40 meV to 100 meV. The angular resolution of the SPECS analyser was 0.07° and that of the DA30L analyser was 0.1°. Fresh surfaces were obtained by cleaving PdBiSe samples in situ in a vacuum with a pressure of less than $5 \times 10^{-11}$ Torr. The DFT calculations were performed using the projector augmented-wave method [34] and the GGA-PBE exchange-correlation functional [35] as implemented in VASP code [36-38]. 10 electrons ($4d^9 5s^1$) in Pd, 5 electrons ($6s^2 6p^3$) in Bi and 6 electrons ($4s^2 4p^4$) in Se are treated as valence electrons. The kinetic energy cutoff of 500 eV is employed. The Brillouin zones were sampled with a 10×10×10 Γ-centered k-point grids for the bulk calculations.

PdBiSe has a noncentrosymmetric structure [Fig. 1(b)] with space group $P2_13$ (No. 198). The crystal is cubic and chiral, lacking inversion and mirror symmetries. Despite the cubic unit cell, the fourfold rotational symmetry is broken due to the chiral structure. Two important symmetries remaining are the threefold rotation symmetry ($C_3$) along the (111) axis and the twofold screw symmetry along the $z$ and $x$-axis. The calculated electronic band structure in the absence of spin-orbit coupling (SOC) is shown in Fig. 1(d). Without considering the spin degree of freedom, at the Γ point, a band crossing with threefold degeneracy is observed at ~ 0.7 eV below $E_F$, which is protected by the $C_3$ symmetry. Its low-energy quasiparticle excitations can be described by a class-II three-component fermion shown in Fig. 1(a)(ii), whose crossing point is a monopole possessing a topological charge of + 2. On the other hand, at the R point, the bulk bands feature a fourfold degenerate band crossing below $E_F$, which is a charge-2 fourfold fermion with a topological charge of − 2.

With the inclusion of SOC [Fig. 1(e)], due to a lack of inversion symmetry, the bands split at all momenta, commonly known as a Rashba-type band splitting. There are two exceptions in the momentum space where degeneracy remains: (i) time reversal invariant momenta, where Kramers theorem guarantees the double degeneracy, and (ii) Brillouin zone boundaries (X-M and M-R), where the screw symmetry ensures the double degeneracy as well. At certain time-reversal invariant momenta, the bulk bands have different types of symmetry-enforced crossings near $E_F$, which we identify with various topological fermions from detailed first-principle calculations. First, at the Γ point, the crossings are either twofold or fourfold degenerate, corresponding to a Weyl fermion and a Rarita-Schwinger-Weyl fermion with $C = + 4$. Second, at the R point, there is a band crossing with sixfold degeneracy, which is a double class-II three-component fermion with $C = - 4$. Third, the bulk bands at the M point also have a fourfold degenerate points at ~ 1.1 eV below $E_F$, which are charge-2 fourfold fermions with $C = + 2$. Lastly, PdBiSe also hosts type-I and type-II Weyl fermions at the X point and along the Γ-X, Γ- R directions [Figs. 1(e) and 1(f)], respectively.

To test the predictions by our calculations, we first perform core-level photoemission measurements to confirm the chemical composition of PdBiSe [Fig. 1(g)]. Next, we systematically investigate the bulk electronic structure using soft X-ray ARPES measurements on the (001) surface. The high energy of soft X-ray leads to a long escape depth of photoelectrons compared to XUV sources, significantly improving the bulk sensitivity as well as the $k_z$ resolution of the measurement [39]. Figure 2 displays the Fermi surfaces in three different high-symmetry planes. The measured Fermi surfaces in the vertical Γ-M-X-R plane (FS1) exhibit a modulation along the $k_z$ direction with a period of $2\pi/c$ [Fig. 2(b)], where $c$ is the lattice constant, confirming the bulk nature of the detected spectra. The Fermi surfaces at $k_z = 0$ (FS2) and $k_z = \pi$ (FS3) exhibit ring-like features centered at the Γ point [Fig. 2(d)] and the M point [Fig. 2(f)], or rhombic pockets surrounding the R point [Fig. 2(f)], which are in good agreement with calculations [Figs. 2(e) and 2(g)]. Note that the splitting of Fermi surfaces is not resolved in Figs. 2(b), 2(d), and 2(f) under current momentum and energy resolution, however, the band splitting can be resolved by high-precision measurements of the band dispersions, as will discuss later. The consistency between the measured and calculated Fermi surfaces gives us confidence to search for the predicted symmetry-stabilized degenerate crossings at various time reversal invariant points.

In the following, we demonstrate the signatures of different topological fermions in PdBiSe by using high-precision measurements of its band dispersion. We start by showing the topological fermions at the Γ point. We observe a crossing point at ~ 0.85 eV below $E_F$ [Figs. 3(a) and 3(b)], which matches our theoretical prediction of a Rarita-Schwinger-Weyl point protected by the $C_3$ rotational symmetry [Fig. 3(c)]. From the calculation, this crossing is fourfold degenerate, resulting from four electron-like bands along the Γ–X direction, labeled 1 to 4 in Fig. 3(c). As there is no inversion symmetry, the four bands are no longer degenerate away from the Γ or the X point, though the splitting is only significant between bands 1 and 2. This Rashba-type splitting is clearly observed in Figs. 3(a) and 3(b) (orange arrows). At the Γ point, we further observe an electron-like band whose bottom is at ~ 0.7 eV below $E_F$. Based on the calculation in Fig. 3(c), this bottom corresponds to a Weyl point protected by the Kramers theorem. Furthermore, we observe a band crossing at ~ 0.25 eV below $E_F$ at the X point [Figs. 3(a) and 3(b)], around which there exist multiple Weyl points, though the fine features of them is not resolved [Fig. 1(f)].

Next, we reveal the evidence of a double class-II three-component fermion with sixfold band degeneracy at the R point. To visualize the degenerate crossing, we map out the band dispersions along the M-R direction with two photon energies that correspond to $k_z = 27\pi$ and $29\pi$. The measured electronic structure exhibits four doubly-degenerate bands around the R point, labeled 1 to 4 in Fig. 3(e). Among them, bands 1 to 3 are degenerate at R, forming a sixfold degenerate crossing, fully consistent with the theoretical calculations [Fig. 3(f)].

Last, to show the existence of charge-2 fourfold fermions at M point, we examine the M-X dispersion along the Brillouin zone boundary [Figs. 3(g)-3(i)]. Due to the screw symmetry, each band in this direction is twofold degenerate. In Figs. 3(g) and 3(h), we resolve two doubly degenerate dispersions, which meet at the M point at ~ 1.1 eV below $E_F$. The observed bands closely follow the calculated structure [Fig. 3(i)], demonstrating the existence of a charge-2 fourfold fermion at the M point.

One hallmark of topological fermions in solids is the existence of surface Fermi arcs connecting the projection of monopoles with opposite topological charges. For example, the surface Fermi arcs connecting the projection of Weyl points in TaAs is clearly resolved [15]. However, Fermi arcs associated with the topological fermions in Fig. 3 were not observed in the present experiment for the following reason: Unlike type-I Weyl points in TaAs [12,13], the observed multiple band crossings in Fig. 3 arise from Rashba-type splitting of the bulk bands, which are highly tilted and eventually disperse in the same directions away from the monopoles. Therefore, the projected topological monopoles and the associated surface states are fully covered by the surface projections of these bulk Rashba cones, rendering Fermi arcs indistinguishable in the photoemission measurement [28].

In contrast to the previous well studied Dirac and Weyl fermion systems which are driven by band inversion based on specific material parameters such as lattice constants and strength of SOC, the observed topological fermions in the present work are stabilized by the crystal symmetry and are therefore universal feature of a chiral crystal with space group $P2_13$ (No. 198). In this aspect, our work would open new possibilities to realize various types of topological fermions in many more material candidates with similar symmetry properties. We further note that PdBiSe is a non-centrosymmetric superconductor [40,41]. Hence, its split Fermi surfaces near the $E_F$ are expected to host rare coexistence of spin singlet and spin triplet superconducting pairing states (see Supplemental Material [42] for details). The latter may give rise to Majorana fermions [43,44], enabling nontrivial applications of PdBiSe such as in quantum computers [45,46].

In summary, by using density functional theory calculations and bulk-sensitive soft X-ray ARPES, we predicted and proved the coexistence of four different types of topological fermions in the electronic structure of PdBiSe. We identified twofold Weyl fermions ($C = \pm 1$) at the $\Gamma$ and X point, fourfold Rarita-Schwinger-Weyl fermion ($C = +4$) at the $\Gamma$ point, sixfold double class-II three-component fermion ($C = -4$) at the R point, and charge-2 fourfold fermion ($C = +2$) at the M point. These multiple topological fermions, which could couple through Coulomb interactions, are stabilized at different time-reversal invariant momenta by crystalline symmetries. Our observation thus paves the way for investigating novel physics related to these unconventional fermions as well as their interplay in a single condensed-matter system.

*Note added:* After completing the manuscript, we became aware of several related works posted on arXiv:1809.01312 [47], arXiv:1812.03310 [48], arXiv:1812.04466 [49], and arXiv:1901.03358 [50], showing the observation of class-II three-component fermions ($C = +2$), charge-2 fourfold fermions ($C = -2$) in the CoSi family. In the present work, PdBiSe represents a richer platform with the coexistence of four different types of topological fermions, i.e., Weyl ($C = \pm 1$), Rarita-Schwinger-Weyl ($C = +4$), double class-II three-component ($C = -4$), and charge-2 fourfold fermions ($C = +2$). More importantly, PdBiSe is the first system where topological fermions with monopole charges of $\pm 4$ are experimentally identified. These fermions are topologically distinct from those in the CoSi family.

## Acknowledgements

This work was supported by the Ministry of Science and Technology of China (2016YFA0401000, 2016YFA0300600, 2015CB921300, and 2017YFA0302901), the Chinese Academy of Sciences (XDB28000000, XDB07000000, and QYZDB-SSW-SLH043), the National Natural Science Foundation of China (11622435, U1832202, 11474340, 11474330, 11604273, and 11774399), and the Beijing Municipal Science and Technology Commission (No. Z171100002017018). N.G., B.Q.L. and A.Z. acknowledge support from the National Science Foundation under Grant No. NSF DMR-1809815 (data analysis), and the Gordon and Betty Moore Foundation's EPiQS Initiative grant GBMF4540 (manuscript writing). J.Z.Z. was also supported by the Longshan academic talent research-supporting program of SWUST (17LZX527), and ETH Zurich for funding his visit, A.A.S. acknowledges the support of Swiss National Science Foundation (SNSF) Professorship, NCCR MARVEL and QSIT grants, as well as Microsoft Research. N F.Q.Y. is supported by the DOE Office of Basic Energy Sciences, Division of Materials Sciences and Engineering under award de-sc0010526. Y.B.H. acknowledges support from the CAS Pioneer "Hundred Talents Program" (type C). A.C. acknowledges funding from the Swiss National Science Foundation under the grant No. 200021_165529.

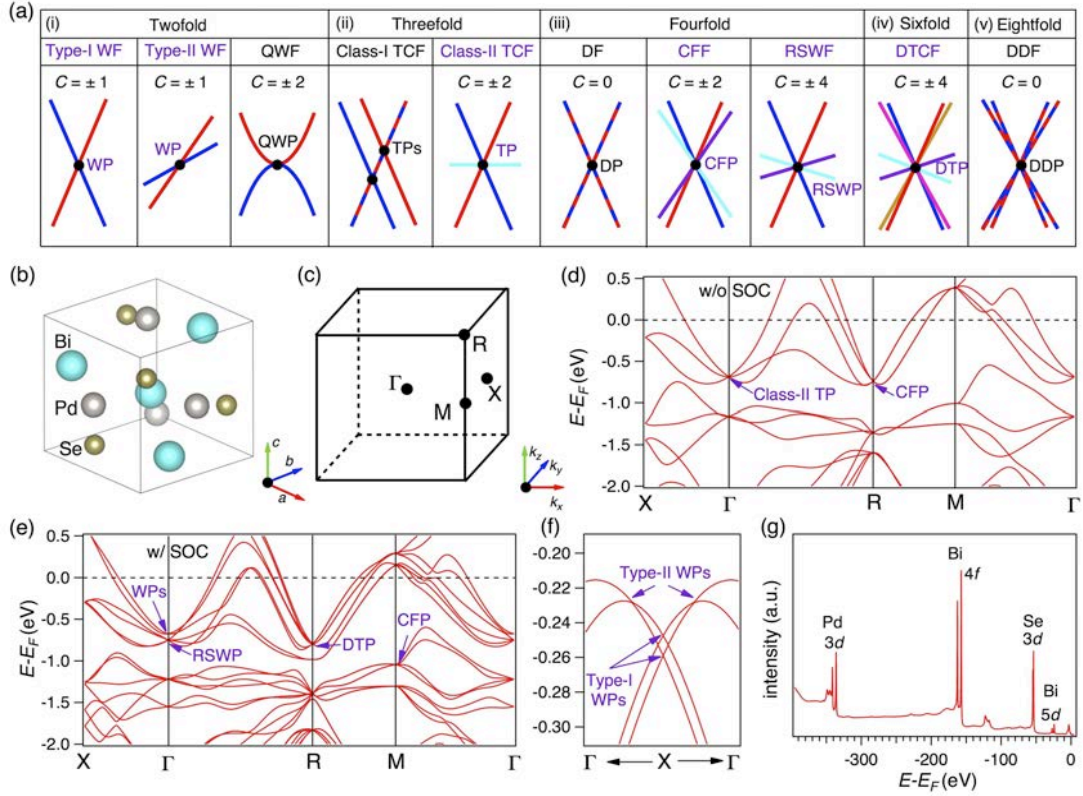

FIG. 1. (a) Schematic of the band structure of various topological fermions in condensed-matter systems. (i) Type-I, type-II and quadratic Weyl fermions with twofold degeneracy, which have topological charges of ± 1, ± 1 and ± 2, respectively. (ii) Class-I three-component fermions with no topological charge and Class-II three-component fermions with a topological charge of ± 2, both of which have band crossings with threefold degeneracies. (iii) Dirac fermion, charge-2 fourfold fermion, and Rarita-Schwinger-Weyl fermion with topological charges 0, ± 2, and ± 4, respectively. All of them have fourfold degenerate band crossings. (iv) Double class-II three-component fermion with sixfold degeneracy. (v) Double Dirac fermion with eightfold degeneracy. (b) 3D crystal structure of PdBiSe. (c) 3D bulk Brillouin zone. (d,e) Calculated band structures of PdBiSe along high-symmetry lines without (d) and with (e) spin-orbit coupling. The purple arrows indicate the momentum locations of multi-band crossings with different band degeneracies and topological charges. (f) Calculated fine band structures, indicating the coexistence of type-I and type-II WPs near the X point with SOC. (g) Core level photoemission spectrum, showing characteristic Pd 3*d*, Bi 4*f*, 5*d* and Se 3*d* peaks. WF/WP: Weyl fermion/point; QWF/QWP: quadratic Weyl fermion/point; TCF: three-component fermion, TP: triple point; DF/DP: Dirac fermion/point; CFF/CFP: charge-2 fourfold fermion/point; RSWF/RSWP: Rarita-Schwinger-Weyl fermion/point; DTCF: double class-II three-component fermion, DTP: double triple point; DDF/DDP: double Dirac fermion/point.

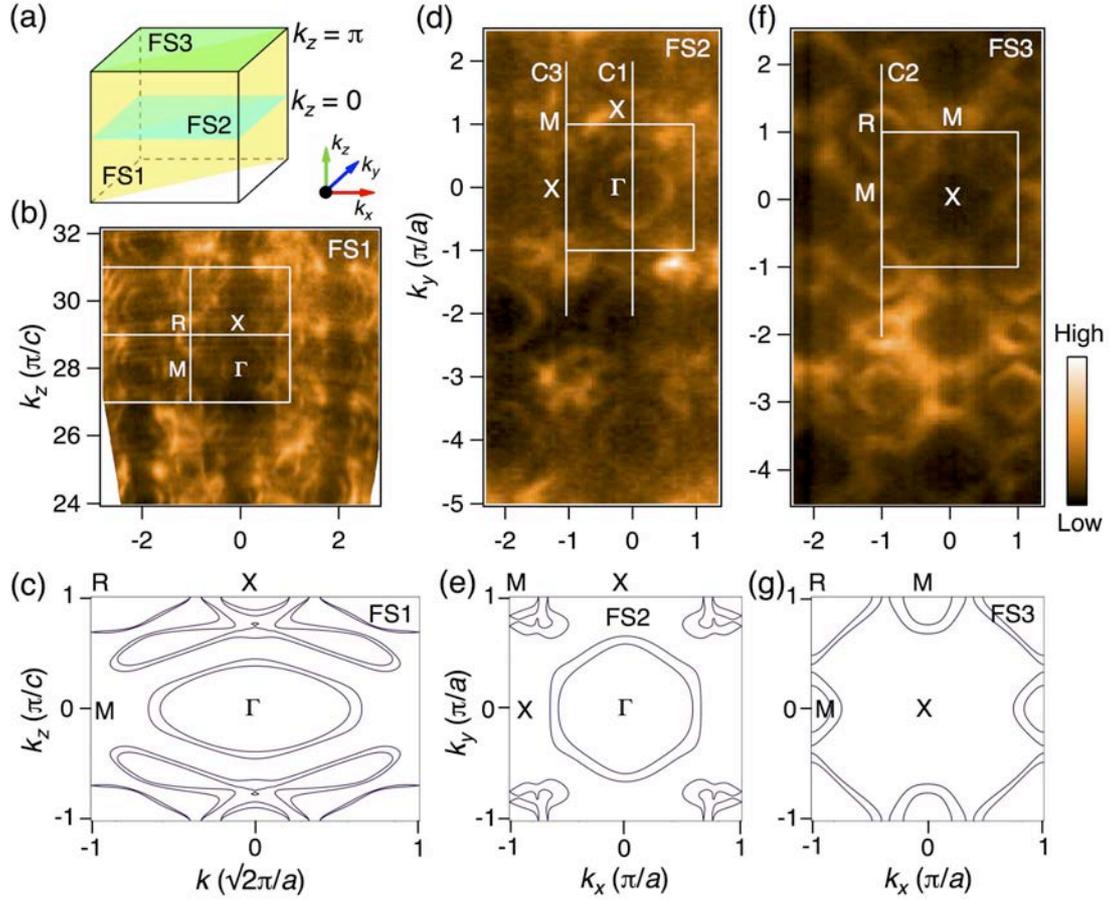

FIG. 2. (a) Bulk Brillouin zone with three high-symmetry planes, which indicate the locations of the measured Fermi surfaces in (b), (d), and (f) respectively. (b) and (c) Experimental (b) and calculated (c) intensity plots at the Fermi level, showing the Fermi surface in the vertical Γ-M-X-R plane (FS1). The white boxes indicate the bulk Brillouin zone boundary. (d) and (e) Experimental (d) and calculated (e) intensity plots at the Fermi level, showing Fermi surfaces in the $k_z = 0$ plane (FS2). The white boxes indicate the Brillouin zone boundary in the $k_z = 0$ plane. (f) and (g) The same as (d) and (e) but in the $k_z = \pi$ plane. The white lines C1, C2 and C3 in (d) and (f) indicate the momentum cuts in Fig. 3. All data were taken on the (001) surface at 20 K.

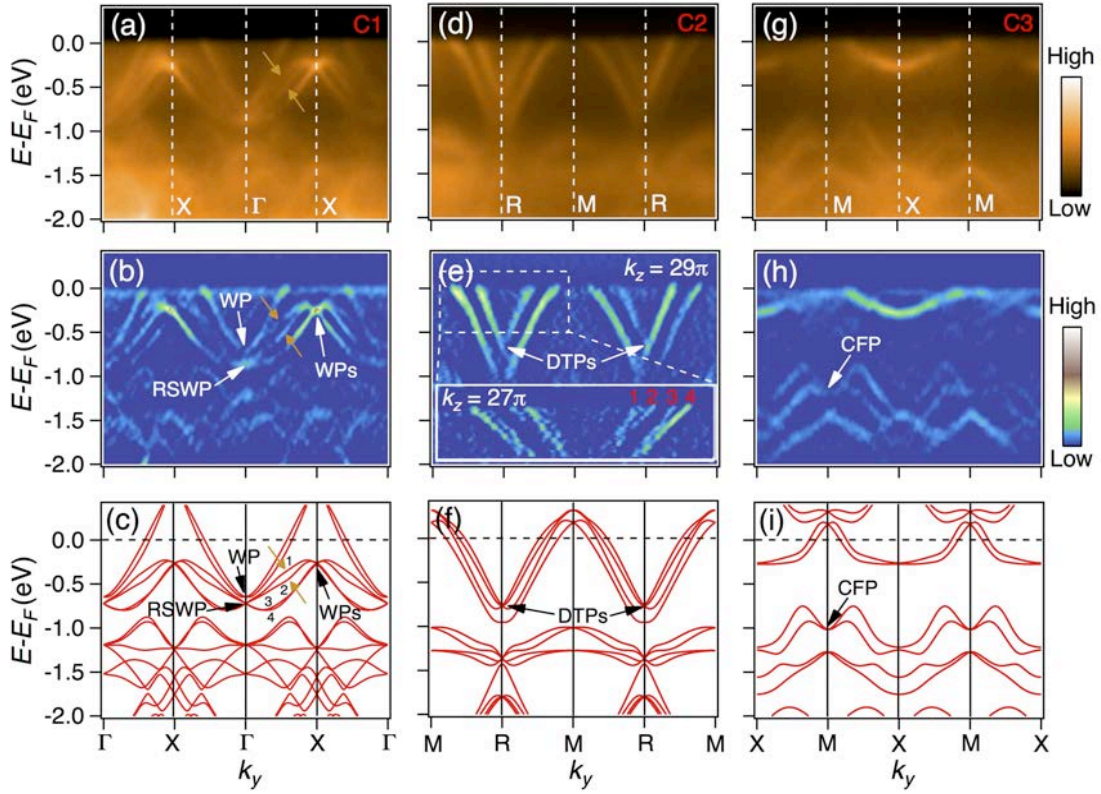

FIG. 3. (a)–(c) ARPES (a) and curvature (b) intensity plots and the calculated band structure (c) along C1 [white line in Fig. 2(d)]. Orange arrows indicate band splitting due to the lack of inversion symmetry. (d)–(f) The same as (a)–(c) but along C2 [white line in Fig. 2(f)]. The normal and enlarged data in (e) were recorded with photo energy $hv$ = 550 eV ($k_z = 27\pi$) and 745 eV ($k_z = 29\pi$), respectively. (g)–(i) The same as (a)–(c) but along C3 [white line in Fig. 2(d)]. The white and black arrows indicate the momentum locations of WPs, RSWPs, DTPs, and CFPs.